\documentclass[preprint,showpacs,preprintnumbers,amsmath,amssymb,nofootinbib]{revtex4}
\usepackage{graphicx}
\usepackage{bm}
\usepackage{epsfig}

\newcommand{\beq}{\begin{equation}}
\newcommand{\eeq}{\vspace{0cm} \end{equation}}
\newcommand{\beqq}{\setlength\arraycolsep{2pt}\begin{eqnarray}}
\newcommand{\eeqq}{\vspace{0cm} \end{eqnarray}}

\begin{document}

\title{On phantom
thermodynamics with negative temperature}

\author{Yun Soo Myung}
  \email{ysmyung@inje.ac.kr}
  \affiliation{Institute of Basic Science and School of
    Computer Aided Science, \\ Inje University, Gimhae 621-749, Korea}
\pacs{Dark energy models, phantom cosmology}

\bigskip
\begin{abstract}
We discuss the thermodynamic properties of the
Friedmann-Robertson-Walker universe with dark energy fluids
labelled by $\omega=p/\rho<-1/3$. Using the integrability
condition, we show that the phantom phase of $w<-1$ can still be
thermodynamically allowed even when the temperature takes on
negative values because in that case, there exists at least a
condition of keeping physical values for $p$ and $\rho$.

\end{abstract}

\maketitle

In the present and future universe, the study of phantom regime is
a very important subject,  since a phantom  fluid of $p+\rho < 0$
violates the strong and dominant energy conditions. Further, the
energy density of a phantom field increases along the cosmic
evolution thereby causing a super accelerating universe which will
end in big rip~\cite{Cald,Barrow}.  The big rip  corresponds to
the case that the singularity of $\rho \rightarrow \infty$ will
appear at a finite time in the future~\cite{NP}.

The thermodynamics of phantom fields is  considered within the
framework of a dark fluid model based on the Euler's relation of
$Ts=(1+\omega)\rho-n\mu$, where $T,s,n,$ and $\mu$ are the
temperature, entropy density,  particle density, and chemical
potential, respectively. In the case of classical fields, there is
no relation to define the particle density $n$ as a function of
scalar field $\phi$ and its derivative $\dot{\phi}$. Hence, we
choose the $\mu=0$ case for simplicity. The phantom phase of
$\omega<-1$ implies simply either $T>0,~s<0$ or $T<0,~s>0$.

In the first approach to phantom thermodynamics, a group of
authors \cite{limamaia,LA04} have insisted that the temperature of
any dark energy component is always positive definite obeying the
evolution law $T \sim a^{-3\omega}$  where $a(t)$ is the scale
factor. In this case, they have shown that the existence of
phantom fluids is not thermodynamically allowed because its
co-moving entropy of $S \sim (1+ \omega)T^{1/\omega}a^{3}$ is
negative.

In the second approach~\cite{gonzalez}, the authors have  claimed
that the temperature  of phantom fluids in the FRW universe is
negative as $T \sim (1+\omega)a^{-3\omega}$ for $\omega<-1$. If
one accepts this temperature reinterpretation, it is possible to
have the positive entropy of the phantom fields.

In this Letter, we derive pressure and energy density as functions
of temperature: $p(T)=C(\omega)T^{(1+\omega)/\omega}$ and
$\rho(T)=\tilde{C}(\omega)T^{(1+\omega)/\omega}$ with unknown
functions $C(\omega)$ and $\tilde{C}(\omega)$. We show that for
$C(\omega)\sim \omega$ and $\tilde{C}(\omega)\sim 1$, the second
approach to phantom thermodynamics with negative temperature is
not thermodynamically allowed since the internal inconsistency
between thermodynamic quantities is arisen from the integrability
condition. However, a specific choice of $C=(1+w)^{2n-(1+w)/w}$
and $ \tilde{C}=C/w$ with $n=0,1,2,\cdots$ leads to a phantom
thermodynamics  because $p(T)/\rho(T)=\omega$ provides a correct
equation of state for $\omega<-1$. This implies that a phantom
thermodynamics with negative temperature is still regarded as a
thermodynamically meaningful system.

We start with the homogeneous and isotropic FRW universe which is
 described by two Friedmann equations based on the
Robertson-Walker metric \beqq  \label{1stf}
H^2&=& \frac{8 \pi G}{3}\rho-\frac{k}{a^2},\\
  \label{secf} \dot{H}&=&-4\pi G(\rho+p)+\frac{k}{a^2} \eeqq
where $H=\dot{a}/a$ is the Hubble parameter and $k=-1,0,1$
represent the three-dimensional space with the negative, zero, and
positive spatial curvature, respectively.

For the thermodynamic study, we apply the combination of the
first- and second-law of thermodynamics to a comoving volume
element of unit coordinate volume and physical volume $V=a^3$.
Then it leads to

 \beq TdS=d(\rho V)+
p dV = d[(\rho+p)V]-V dp \label{kt364}. \eeq
 The  integrability
condition is necessary to define the FRW universe as a
thermodynamic system~\cite{kolb}. It is given by \beq {\partial^2
S \over
\partial T \partial V}={\partial^2 S \over
\partial V \partial T}\label{kt365} \eeq which leads to the
 relation between the pressure (energy density) and temperature  \beq dp={\rho+p\over
T}dT.\label{kt367} \eeq Plugging Eq.(\ref{kt367}) into
Eq.(\ref{kt364}), we have the differential relation, \beq dS =
{1\over T} d[(\rho + p)V]-(\rho + p)V {dT\over
T^2}=d\bigg[{(\rho+p)V\over T} + C \bigg]\label{kt368} \eeq where
$C$ is a constant. The entropy  per comoving volume must be
defined by \beq S \equiv{(\rho + p)\over T}V\label{entropy} \eeq
up to an additive constant.

On the other hand, the conservation law plays an important role in
the FRW universe. This could be derived from Eqs.(\ref{1stf}) and
(\ref{secf}) as \beq \label{con-law} \dot{\rho}+3H(\rho+p)=0 \eeq
 Hence, one
equation among three is redundant. Here we choose  the first
Friedmann equations (\ref{1stf})  and the conservation law
(\ref{con-law}) as the relevant equations. Importantly,
Eq.(\ref{con-law}) gives the energy density with $p=\omega \rho$
\beq \label{rho-a}\rho(a)=\rho_0 a^{-3(1+\omega)}.\eeq Considering
the equivalent relation of $d(\rho V) + pdV=0$, the conservation
law corresponds to $dS=0$. That is,  the dark fluid should expand
adiabatically as \beq d\bigg[{(\rho+p)V\over T}\bigg]=0,
\label{kt370} \eeq which means that the entropy $S$ per comoving
volume is conserved. In other words, the FRW universe satisfying
the conservation law should expand adiabatically.

Even for an adiabatic process of $S={\rm const}$, the same
definition of entropy follows from the conservation law  which can
be rewritten as \beq
 d[(\rho+p)V]=V dp.\label{kt369} \eeq Inserting the integrability condition Eq.(\ref{kt367}) into Eq.(\ref{kt369}), one
recovers Eq.(\ref{kt370}) immediately.

Using $p=\omega\rho$, we  rewrite the entropy of
Eq.(\ref{entropy}) as \beq S={(1+\omega)\rho V \over T},
\label{kt371} \eeq which defines the entropy  in terms of the
temperature and energy density.  At this stage, we would like
mention a work on the holographic approach to the FRW
universe~\cite{donam}. This author assumed that Eq. (\ref{kt371})
defines the temperature in terms of the constant entropy  as \beq
\label{temp} T={(1+\omega)\rho_0 \over S}a^{-3\omega}. \eeq At
that time, one implicitly believed that the scaling law $T \sim
a^{-3\omega}$ holds for $\omega \ge -1$ because the literatures
mainly focused on the holographic description for a
radiation-dominated universe with $\omega=1/3$~\cite{frw}.   Later
on, Eq.(\ref{temp}) was used for giving the origin to the idea of
negative temperature for $\omega <-1$ and $S>0$~\cite{gonzalez}.
Since then, an extension to $\omega<-1$ has been widely considered
in many different contexts~\cite{PL}.

However, we wish to show that  this extension  may induce a main
flaw when considering an evolution of the FRW universe with dark
fluid as an adiabatic process of thermodynamic system. From the
integrability condition Eq.(\ref{kt367}), we have \beq
\frac{dp}{p}=\Big[\frac{1+\omega}{\omega}\Big] \frac{dT}{T} \eeq
which leads to \beq \ln p=\Big[\frac{1+\omega}{\omega}\Big] \ln
T+\ln C(\omega). \eeq In this case, we obtain the relation between
pressure and temperature\footnote{ As a simple example, we
consider an ideal gas with constant heat capacities undergoing a
reversible, adiabatic compression and expansion.  In this case, we
have an adiabatic process $dU+pdV=0$ with $dU=C_VdT$, which leads
to $C_VdT=-pdV$. Using its equation of state  $pV=RT$ which
implies $dV=-\frac{RT}{p^2}dp+\frac{R}{p}dT$, one finds
$C_VdT=RT\frac{dp}{p}-RdT$. Considering $C_V+R=C_p$, it leads to
$C_p dT=RT \frac{dp}{p}$ which can be rewritten as
$\frac{dp}{p}=\Big[\frac{\gamma}{\gamma-1}\Big]\frac{dT}{T}$ with
$\gamma=C_p/C_V$. Integration gives a similar formula for an
adiabatic process of an ideal gas as
$p=c(\gamma)T^{\frac{\gamma}{\gamma-1}}$.}

 \beq
p(T)=C(\omega)T^{\frac{1+\omega}{\omega}} \label{pt-law}\eeq with
an unknown function $C(\omega)$.
 Also, from the other
form of integrability condition \beq
\frac{d\rho}{\rho}=\Big[\frac{1+\omega}{\omega}\Big] \frac{dT}{T}
\eeq we derive the generalized Stefan-Boltzmann
law\footnote{Actually, this expression differs from
$\rho=\rho_0\Big(T/\kappa(1+\omega)\Big)^{\frac{1+\omega}{\omega}}
$ in~\cite{gonzalez}, which was obtained by plugging
Eq.(\ref{temp}) into Eq.(\ref{rho-a}). In the case of $\omega<-1$,
they must take $T<0$ to preserve $\rho$ positive.}~\cite{LA04}
\beq \label{sb-law} \rho(T)=\tilde{C}(\omega)
T^{\frac{1+\omega}{\omega}} \eeq with an unknown function
$\tilde{C}(\omega)$.
 At the first glance, it seems
that one recovers the important  relation of $p \sim \omega \rho$
if $C(\omega)\sim \omega $ and $\tilde{C}(\omega) \sim 1$. Then,
substituting $T$ in Eq.(\ref{temp})  into Eq.(\ref{pt-law}) leads
to

\beq p(T) \label{pteq}
    =\omega
    (1+\omega)^{\frac{1+\omega}{\omega}}\Big[\frac{\rho_0}{S}\Big]^{\frac{1+\omega}{\omega}}a^{-3(1+ \omega)}.\eeq
Similarly, plugging $T$ into Eq.(\ref{sb-law}) leads to \beq
\rho(T)=
    (1+\omega)^{\frac{1+\omega}{\omega}}\Big[\frac{\rho_0}{S}\Big]^{\frac{1+\omega}{\omega}}a^{-3(1+ \omega)}.\eeq
\begin{figure}[t!]
   \centering
   \includegraphics{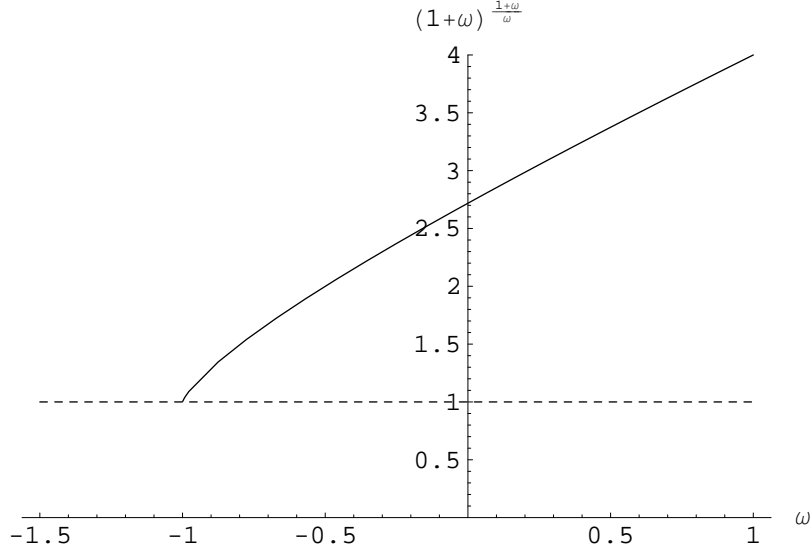}
\caption{Graph of $(1+\omega)^{\frac{1+\omega}{\omega}}$ as
function of $\omega$. The dotted line denotes ``1". } \label{fig1}
\end{figure}
Here we observe that if $\omega<-1$ and $S>0$, these two
expressions are not defined properly because of the term  of
$(1+\omega)^{\frac{1+\omega}{\omega}}$. For example, we have
$(-0.1)^{1/11}$ for $\omega=-1.1$, which does not provide a real
value. As is shown Fig. 1, $(1+\omega)^{\frac{1+\omega}{\omega}}$
is not available  for $\omega<-1$ and it has a limiting value 1
for the vacuum state of $\omega=-1$. This arises because we
introduce the first- and second-law with the integrability
condition to describe the FRW universe with dark fluid in terms of
a thermodynamic system, in addition the conservation law. We note
that for the phantom regime,  the conservation law of
Eq.(\ref{con-law}) is not compatible with the integrability
condition of Eqs.(\ref{pt-law}) and (\ref{sb-law}).

It is suggested that if the temperature of a dark energy fluid is
negative,  the phantom regime ($\omega <-1$) is thermodynamically
forbidden, even though  its entropy is positive.

However, we note that  this statement may be changed when one
chooses specific forms  of $C(\omega)$ and $\tilde{C}(\omega)$.
For example, one may take $C=(1+w)^{2n-(1+w)/w}$ and $
\tilde{C}=C/w$ with  $n=0,1,2,\cdots$\footnote{The author thanks
anonymous referee for pointing out this case.}. It seems to appear
a little more reasonable because as one approaches the big rip,
one would expect a given quantization characterized by $n$ to take
place. In this case, a phantom phase may be allowed because
$p(T)/\rho(T)=\omega$ provides a correct equation of state for
$\omega<-1$.

The vacuum state of $\omega=-1$ corresponds to the marginal  case:
The total energy  $E=\rho V$  increases during the expansion,
while its energy density is constant~\cite{LA04}. In this case, if
 entropy is zero, the temperature is ill-defined. The reverse case
 is also possible to occur. This means that the thermodynamic
 interpretation for the vacuum state is still obscure.

Finally, one may introduce  a negative chemical potential to avoid
negative temperature~\cite{PL,Per1}. In this case, both the
entropy and temperature of the phantom fluid may be positive when
the chemical potential is negative. However,  there exist
controversial issues to find the form of chemical potential for
the FRW universe~\cite{Pach2} and the accretion of phantom fields
by black holes~\cite{PH,LPHG}.

\begin{acknowledgments}
This was in part supported by the Korea Research Foundation
(KRF-2006-311-C00249) funded by the Korea Government (MOEHRD) and
the SRC Program of the KOSEF through the Center for Quantum
Spacetime (CQUeST) of Sogang University with grant number
R11-2005-021.

\end{acknowledgments}

\end{document}